\documentclass[final,1p,times]{elsarticle} 

\usepackage{graphicx}
\usepackage{amssymb} 
\usepackage{amsthm} 
\usepackage{lineno}

\journal{Nuclear Physics A} 

\begin{document}

\begin{frontmatter} 

 \title{Isolated photon-hadron correlations in proton-proton collisions at $\sqrt{s}$ = 7 TeV with the ALICE experiment}

\author{Nicolas Arbor (for the ALICE Collaboration)}

\address{LPSC, Universit\'e Joseph Fourier Grenoble 1, CNRS/IN2P3, Institut Polytechnique de Grenoble, 53 rue des Martyrs, Grenoble, France}

\begin{abstract} 
At high $p_T$, direct photons produced in Compton and annihilation QCD leading order processes are associated to a jet in the opposite direction. Such processes are tagged experimentally by identifying leading isolated photons and their correlated associated hadrons in the opposite azimuthal direction. The jet fragmentation can be estimated from the hadrons and the photon via the imbalance parameter $x_E = -\vec{p}_T^h.\vec{p}_T^\gamma / |\vec{p}_T^\gamma|^2$. We present the results extracted from gamma-hadron correlations measured by the ALICE experiment in pp collisions at $\sqrt{s}$ = 7 TeV. Direct photons are first identified using isolated criteria. Then, the remaining contamination from neutral mesons decay photons is subtracted statistically to extract the $x_E$ distributions of isolated photon-hadron and isolated $\pi^0$-hadron correlations.   
\end{abstract} 

\end{frontmatter} 


High-$p_T$ partons and photons produced in hard processes are often used as probes to study the strongly interacting medium created in heavy-ion collisions. As the production of these energetic particles occurs slightly before the creation of the medium, high-$p_{T}$ partons can traverse the medium before they hadronize in a jet of hadrons. The modification of the parton fragmentation resulting from the energy loss (via collisional or radiative mechanisms) along the medium path can be used to infer medium properties. At the Large Hadron Collider (LHC), medium effects were first studied using back-to-back dijets or dihadron correlations \cite{1}\cite{2}. However correlating two probes that both interact with the medium induces a surface bias in which the sampled hard scatterings are likely to occur at, and to be oriented tangential to, the surface of the medium \cite{PHENIX}. In contrast to partons, direct photons are not colored objects and hence escape almost unmodified from the medium. At leading order (LO), their production in pp collisions is dominated by $qg$ Compton scattering and $q\bar{q}$ annihilation. Photons emitted from such processes may be used to estimate the initial transverse momentum of the recoil parton. These tomographic studies of the medium using parton energy loss in heavy-ion collisions requires first detailed measurement of direct photon-hadron correlations in pp collisions.

The measurement presented here consists in selecting direct photons in coincidence with charged hadrons to access the away-side parton fragmentation. The fragmentation function should be given to a good approximation by the $x_E$ distribution where,
\begin{center}
{\large $x_{E} = -\frac{\vec{p}_T^h.\vec{p}_T^\gamma}{|\vec{p}_T^\gamma|^2} = -\frac{|p_T^h|cos\Delta\Phi}{|p_T^\gamma|}$},
\end{center}
with $\Delta\Phi$ corresponding to the azimuthal angle between photons and hadrons. Figure 1 illustrates the $x_E$ distribution computed from a Diphox $\gamma$-jet production \cite{3} and compared with the DSS quark and gluon fragmentation functions \cite{DSS}. It shows that even if the $x_E$ variable is not an exact measurement of the fragmentation function because of higher-order effects, the $x_E$ distribution follows the fragmentation behaviour over a large range (mainly the quark fragmentation due to the dominating contribution of the Compton scattering cross-section).

\begin{figure}[htbp]
\begin{center}
\includegraphics[,height=5.9cm,width=0.49\textwidth]{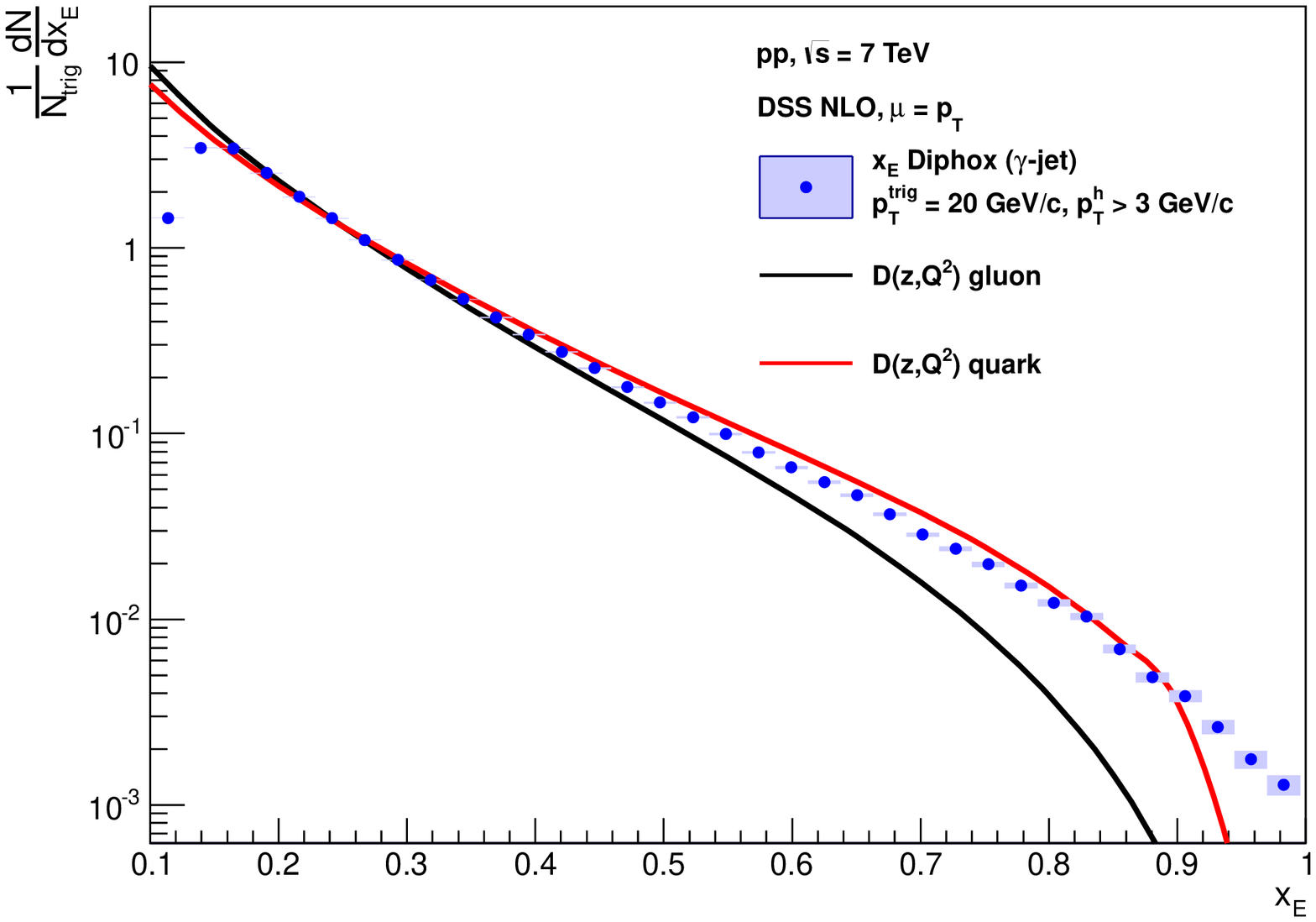}
\includegraphics[,height=6cm,width=0.49\textwidth]{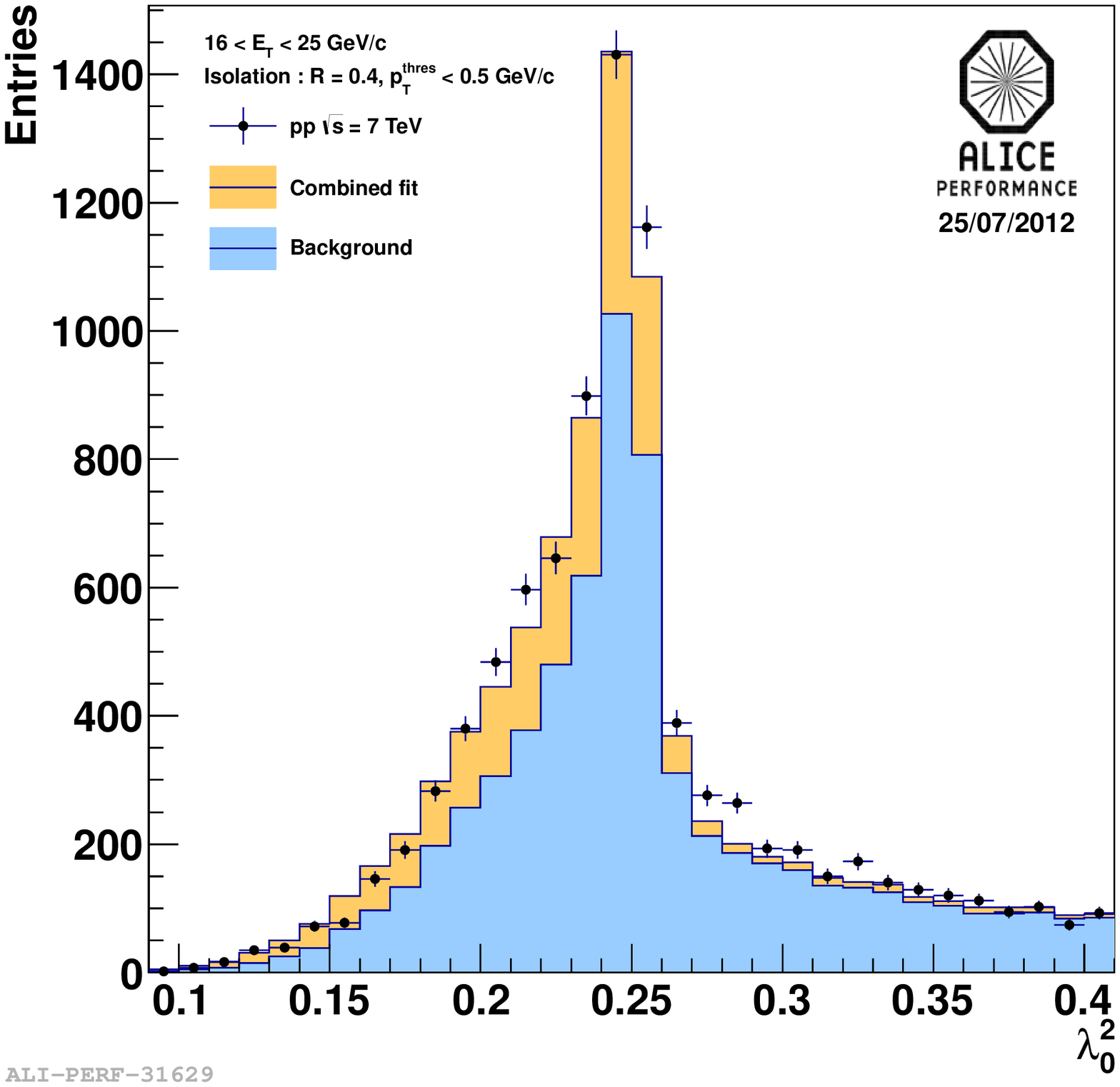}
\end{center}
\caption{Left: $\gamma$-hadron $x_E$ distribution compared to DSS quark and gluon fragmentation functions (arbitrary scaled) ; Right: Isolated clusters shower shape long axis distribution fitted with a two-component binned likelihood.}
\label{fig:diphox}
\end{figure}

This analysis uses approximately 10 million events from pp collisions at $\sqrt{s}$ = 7 TeV ($L_{int} \approx$ 500 $nb^{-1}$) recorded by the ALICE detector in 2011 \cite{ALICE}. Photons are detected in the EMCal electromagnetic calorimeter which is a Pb-Scintillator sampling electromagnetic calorimeter that covers $\Delta\Phi = 100^\circ$ in the azimuthal angle and $|\eta| < 0.7$ in pseudo-rapidity, with a granularity of $\Delta\eta$ $\times$ $\Delta\Phi$ = 0.0143$\times$0.0143. The EMCal level 0 trigger was used, applying a 5 GeV/c threshold, to achieve the measurement of high-$p_T$ photons up to 25 GeV/c with enough rate. Photon candidates are reconstructed from energy clusters deposited in the $p_T$ range from 8 to 25 GeV/c. On the opposite side, charged hadrons are detected with the ALICE tracking system in the central barrel which consists of a silicon detector (Inner Tracking System) and a time projection chamber (TPC). The ITS consists of six layers equipped with silicon pixel detectors (SPD), silicon drift detectors (SDD) and silicon strip detectors (SSD). The two innermost layers cover a pseudo-rapidity range of $|\eta| < 2$ and $|\eta| < 1.4$, respectively. The TPC is a cylindrical drift detector with an acceptance of $|\eta| < 0.9$ in the full azimuthal angle. Data are corrected for detector effects, such as tracking efficiency and energy resolution, to avoid biasing the fragmentation measurement both on the trigger (photon) and on the away (hadrons) sides. 

The single clusters distribution is dominated by a large background coming from electromagnetic decays of neutral mesons (mostly $\pi^0$). This background can be reduced by about 80\% by imposing isolation criteria on the photon candidates, based on the fact that neutral mesons are produced inside a jet and are surrounded by hadronic activity. In this analysis, isolation criteria are based on the absence of particles with a $p_T > 0.5$ GeV/c in a cone of radius $R = \sqrt{\Delta\eta^2+\Delta\Phi^2}$ = 0.4 around the photon candidate. The isolation also suppresses a significant fraction of the fragmentation photon component. To construct photon-hadron correlations, the leading isolated photon is associated to every charged hadron on the opposite hemisphere of the event with a $p_T >$ 0.2 GeV/c. Background contributions coming from isolated meson decays that carry a large fraction of the parent parton need to be subtracted. The shape of the electromagnetic shower produced inside the calorimeter may be used both to reject decay photons and to estimate the remaining contamination in the isolated particles sample. Photons interacting with the calorimeter produce an electromagnetic shower that deposits energy in several cells. Those cells are grouped forming a cluster using a simple algorithm that aggregates all cells with common edges if they have more than 50 MeV. The energy redistribution in the cells has a shape that can be used to distinguish single photons from other particles and in particular from $\pi^0$ decay photons.  The cluster shape is quantified using the long axis of the ellipsoidal parametrization of the energy deposit matrix :
\begin{center}
$\lambda_0^2 = 0.5\times(d_{\eta\eta}+d_{\phi\phi})+\sqrt{0.25\times(d_{\eta\eta}-d_{\phi\phi})^2+d_{\eta\phi}^2}$,
\end{center}
with $d_{ii}$ corresponding to energy distribution in the direction $i$. As shown in Fig.1 (right), the long axis length, $\lambda_0$, is fitted by the signal and background distributions using a two-component binned likelihood. The signal component is obtained from $\gamma$-jet events generated with PYTHIA and propagated through the detectors with GEANT3. The background component shape (mainly $\pi^0$) is extracted from data by taking the shower shape distribution of events which have failed the isolation criteria. Typical purity values extracted from fit results in the 8 to 25 GeV/c $p_T$ range go from about 5 to 70\%.

\begin{figure}[htbp]
\begin{center}
\includegraphics[width=0.49\textwidth]{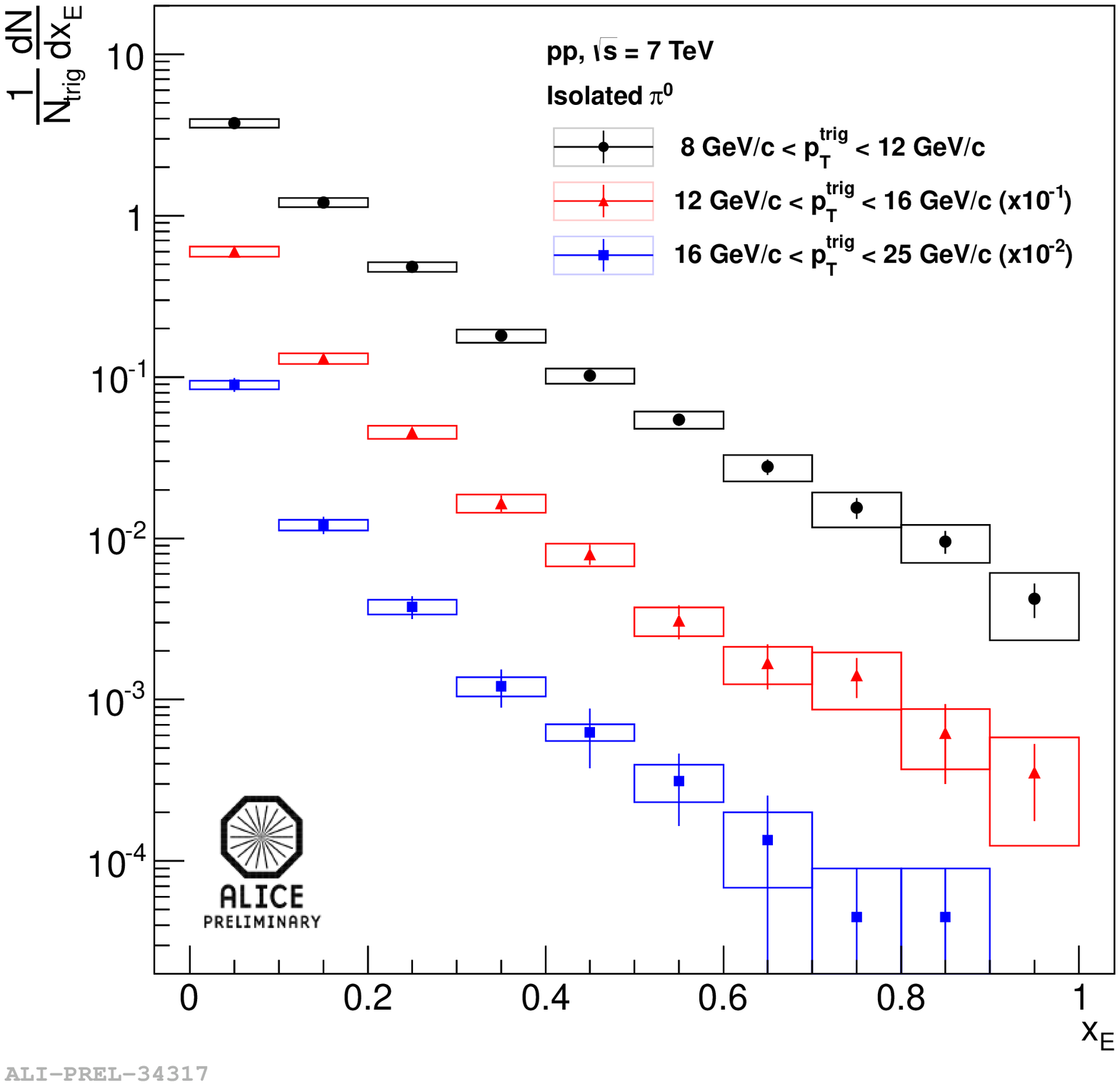}
\includegraphics[width=0.49\textwidth]{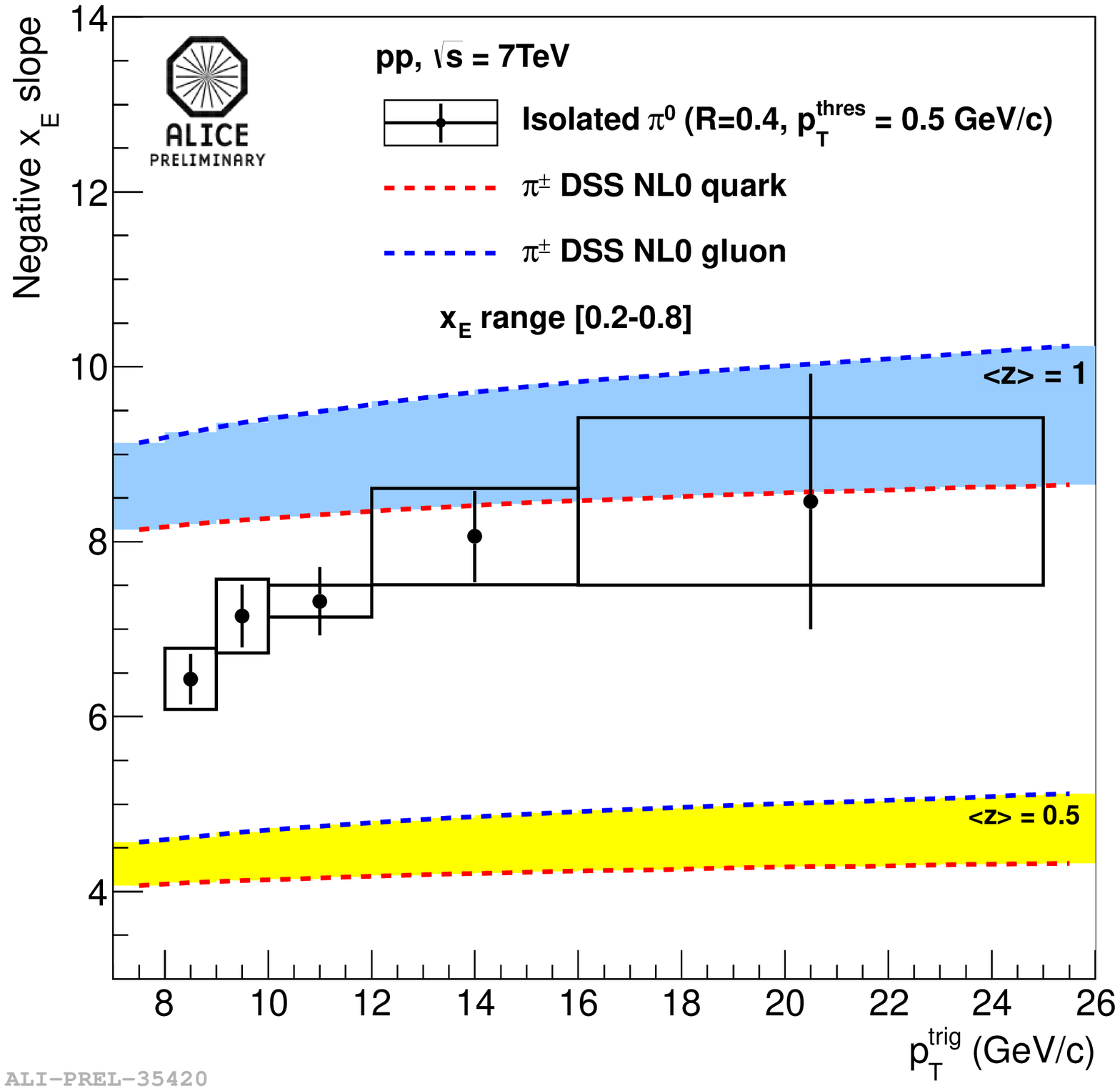}
\end{center}
\caption{Left: $x_E$ distributions of $\pi^0$-hadron correlations for 3 $p_T$ bins ; Right: slopes extracted from exponential fit of isolated $\pi^0$-hadron correlations (5 $p_T$ bins) and compared to DSS quark-gluons fragmentation functions.}
\label{fig:xE}
\end{figure}

The $x_E$ distribution is obtained by selecting hadrons within $2\pi/3 < \Delta\Phi < 4\pi/3$. To subtract the contamination coming from neutral meson decays, we have used $x_E$ distribution of isolated $\pi^0$-hadron correlations and scaled this distribution with respect to the purity estimate described above. The isolated $\pi^0$-hadron correlations results are shown for three $p_T$ bins of the triggered $\pi^0$ in Fig.2 (left). Isolated photon-hadron correlations are then corrected for the soft background coming from correlations between photon and hadrons not originating from the hard scattering such as initial/final-state-radiations or multiple parton interactions. This background has been estimated using two different underlying event regions $\pi/3 < \Delta\Phi < 2\pi/3$ and $4\pi/3 < \Delta\Phi < 5\pi/3$. The resulting $x_E$ distribution of isolated photon-hadron correlations has been extracted for a $p_T$ range from 8 to 25 GeV/c. An exponential slope can then be obtained by fitting the final $x_E$ distribution (Fig.3). In addition to the isolated photon-hadron correlations analysis, we have studied the possibility of using the isolated $\pi^0$ as a trigger particle. The isolation selects mainly $\pi^0$ which carry a large fraction of the total jet energy. Thus, it increases significantly the partonic momentum fraction $<z> = p_T^{\pi^0}/p_T^{parton}$ carried by the $\pi^0$ compared to the expected value $<z> \approx 0.5$ without isolation. However, a slope comparison between isolated $\pi^0$-hadron $x_E$ distributions and DSS fragmentation functions (fig.2 right) shows that isolated $\pi^0$-hadron correlations still suffer from the fact that the $\pi^0$ is itself a parton fragment with $p_T^{\pi^0} < p_T^{parton}$.  

Isolated photon-hadron correlations are one of the most promising channels to study the parton energy loss. The detailed analysis done by the ALICE experiment on the 2011 pp data is an important step toward a better comprehension of the medium modified fragmentation function. The slope parameter, and the $x_E$ distribution itself, form an essential baseline for the on-going Pb-Pb collisions analysis.

\begin{figure}[htbp]
\begin{center}
\includegraphics[width=0.49\textwidth]{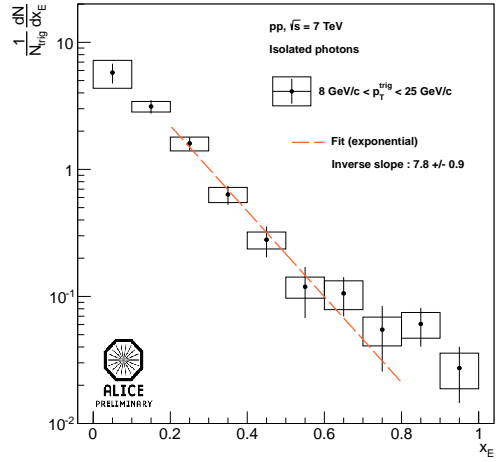}
\end{center}
\caption{$x_E$ distribution of isolated $\gamma$-hadron correlations fitted by an exponential in the $x_E$ range [0.2-0.8].}
\label{fig:xE2}
\end{figure}

\end{document}